\documentclass[journal]{IEEEtran}
\usepackage{cite}

\ifCLASSINFOpdf
  \usepackage[pdftex]{graphicx}
\else
\fi
\usepackage{amsmath}
\usepackage{physics}
\usepackage{bbold}
\usepackage{algorithmic}
\usepackage{algorithm}

\usepackage{array}
\newcolumntype{P}[1]{>{\centering\arraybackslash}p{#1}}
\newcolumntype{M}[1]{>{\centering\arraybackslash}m{#1}}

\usepackage{color}

\usepackage{fixltx2e}
\begin{document}
%
\title{Energy Efficiency and Hover Time Optimization in UAV-based HetNets}
%
%
%

\author{Sidra~Tul Muntaha,
        Syed Ali~Hassan,~\IEEEmembership{Senior Member,~IEEE,}\\
        ~Haejoon Jung,~\IEEEmembership{Member,~IEEE,} 
        ~and M. Shamim Hossain,~\IEEEmembership{Senior Member,~IEEE}
\thanks{
This work was supported in part by the Basic Science Research Program through the National Research
Foundation of Korea (NRF) funded by the Korea Government (Grant Number NRF-2019R1F1A1047989) and in part by the Researchers Supporting Project Number (RSP-2020/32), King Saud University, Riyadh, Saudi Arabia. \emph{(Corresponding author: Haejoon Jung.)}}
\thanks{S. T. Muntaha and S. A. Hassan are with the School of Electrical Engineering and Computer Science, National University of Sciences and Technology, Islamabad 44000, Pakistan (e-mail: smuntaha.msee16seecs@seecs.edu.pk; ali.hassan@seecs.edu.pk).}
\thanks{H. Jung (corresponding author) is with the Department of Information and Telecommunication Engineering, Incheon National University, Yeonsu-gu, Incheon, Korea (e-mail: haejoonjung@inu.ac.kr).}
\thanks{M. S. Hossain is with the Department of Software Engineering, College of Computer and Information Sciences, King Saud University, Riyadh 11543, Saudi Arabia (e-mail: mshossain@ksu.edu.sa).}}

\maketitle

\begin{abstract}
In this paper, we investigate the downlink performance of a three-tier heterogeneous network (HetNet). The objective is to enhance the edge capacity of a macro cell by deploying unmanned aerial vehicles (UAVs) as flying base stations and small cells (SCs) for improving the capacity of indoor users in scenarios such as temporary hotspot regions or during disaster situations where the terrestrial network is either insufficient or out of service. UAVs are energy-constrained devices with a limited flight time, therefore, we formulate a two layer optimization scheme, where we first optimize the power consumption of each tier for enhancing the system energy efficiency (EE) under a minimum quality-of-service (QoS) requirement, which is followed by optimizing the average hover time of UAVs. We obtain the solution to these nonlinear constrained optimization problems by first utilizing the Lagrange multipliers method and then implementing a sub-gradient approach for obtaining convergence. The results show that through optimal power allocation, the system EE improves significantly in comparison to when maximum power is allocated to users (ground cellular users or connected vehicles). The hover time optimization results in increased flight time of UAVs thus providing service for longer durations.
\end{abstract}


%
\IEEEpeerreviewmaketitle

\section{Introduction}
%
%
%
%
\IEEEPARstart{O}{ver} the past few years, the demand for higher data rates and uninterrupted data communication has radically increased. With the newer applications of Internet of things (IoT) coming to light such as Internet of connected vehicles (IoCV) and vehicle to everything (V2X) communication, it is anticipated to have an exponential growth in the number of data users and connected vehicles. This amplification in user density requires a scalable and dynamic network. As a result of latest research and development in fifth generation (5G) and beyond 5G (B5G) technology, forming heterogeneous networks (HetNets) is one way to cater for such escalated demands. In 5G and B5G, the concept of unmanned aerial vehicles (UAVs) have become very popular as they provide cost-efficient and easy to deploy solutions for improving data rates of ground cellular users and connected vehicles. UAVs have vast applications in scenarios such as provision of data service to hotspot regions such as football grounds, open-air events, traffic jams etc., in certain disaster situations such as earthquakes and Tsunamis especially when the terrestrial network is no longer in service, and to enhance network coverage of cell edge users. \color{black} {Features such as mobility and varying altitude make it easier for UAVs to form better line-of-sight (LoS) connections with the users and provide coverage to a wider region as compared to small cells (SCs) which are low-powered base stations that provide high data rates for indoor users over a small region. In contrast to UAVs, dense deployment of SCs comes at the cost of increased interference among the users. On the other hand, a single UAV can replace many small cells while maintaining the same throughput thus significantly decreasing co-tier interference caused by ultra-dense deployment of small cell base stations \cite{r15,r24}}. \color{black} Forming a HetNet consisting of macro base station (MBS), UAVs, and SCs could be an efficient approach to cater for the rapidly rising data demands posed by users and connected vehicles, enhance system energy efficiency and extend the network coverage \cite{r1,r2,r3,r4,r5,r6,r7,r8}. In the following subsection, we briefly highlight the related work in this area.

\subsection{Literature Review}
In \cite{r9}, the authors have taken into consideration the power efficient deployment of UAVs while satisfying the users’ rate requirement. The authors in \cite{r10,r11} deploy UAVs as flying base stations (FBS) to enhance the coverage area and data capacity of a macrocell under minimum power requirements. The authors in \cite{r12,r13} deals with the 3-D placement of UAVs to enhance network capacity of terrestrial networks. UAVs are typically designed for servicing outdoor users, but in \cite{r25}, UAVs are placed in a manner to provide coverage to the indoor users in a high rise building when the terrestrial network is out of service. The traditional channel modeling for terrestrial networks is not suitable for UAVs because of factors such as mobility, varying altitude, and distinctive power constraints. Therefore, air-to-ground (A2G) channel models are formed for the operation of UAVs. The authors in \cite{r14} describe the factors that affect the A2G channel model. \color{black}{The probability of having LoS and NLoS signals depends upon the environment, small-scale and large-scale fading, and most importantly height and placement of a UAV}. 

\color{black} The authors in \cite{r15} discuss the performance of UAVs when they are used as FBS, which are proven to provide enhanced performance as compared to ultra-dense fixed small base stations (SBSs). The authors have considered the mobility of users as well as the variation in their data requirement to get more realistic results, which makes the FBS a far better choice than static base stations. In \cite{r16}, the authors investigate the effect of offloading user data from ground base stations to UAVs in case of crowded situations and show that the overall data delivery improves this way. In \cite{r17}, the UAVs acting as FBS are repositioned so that the spectral efficiency of the network increases. When UAVs are repositioned according to varying user locations, there are better chances of LoS connections and fewer packet losses. In \cite{r2,r19}, the hover time of UAVs is optimized through efficient cell partitioning of UAVs in an all-UAV network. Cells are formed on the basis of user density, therefore, a user fairness is obtained. The area with higher user density gets smaller partition and vice versa. This way the UAVs hover on their respective partitions for almost equal times. \color{black} In \cite{r7}, UAVs are deployed as FBS with the purpose of collecting data from the IoT devices. This kind of network requires an energy efficient system so that UAVs can operate for longer durations. The authors in \cite{r7} have adopted several techniques to make this an energy efficient system by optimizing the deployment, mobility, trajectory and power of UAVs. Because of optimal placement and mobility of UAVs, the UAVs as well as the IoT devices consume much less power. \color{black}

\color{black} The authors in \cite{r5} have considered a HetNet comprising of three tiers, i.e., macro cell, SBSs and UAVs. Ultra-high frequency (UHF) band is used for macro cell and UAVs, while SBSs utilize the millimeter wave (mmWave) band. The number of UAVs and SBSs are fixed. In that work, the authors have enhanced the energy efficiency of the system by optimizing power.\color{black} In \cite{r18}, UAVs are being used as FBS and the network is made energy efficient by optimizing the trajectory of UAVs. \color{black} The authors in \cite{r20} have proposed dynamic path planning for QoS provision to UAV environments. \color{black} The authors in \cite{r21,r22} investigate a software defined network (SDN)-based approach for ensuring the provision of QoS and secure communication to connected vehicles. UAV networks are prone to attacks which leads to an insecure network. In \cite{r23}, a tree based attack-defense model is proposed for risk assessment of such attacks.

To the best of our knowledge, none of the literature reviewed in [11]--[25] has investigated the effects of hover time optimization of UAVs in a three-tier HetNet while optimizing the power consumption of each tier along with managing the interference constraint. Optimizing hover time will allow the UAVs to service the ground users and connected vehicles for longer durations. Taking this constraint into consideration will enhance the overall system efficiency.


\subsection{Contributions}
The main contribution of this paper is to investigate the effects of optimizing the hover time of UAVs in a three-tier HetNet with varying number of UAVs to enhance the cell coverage of a macro cell. In this paper, we adopt a two-layer optimization structure for enhancing the energy efficiency of the system. Firstly, the power consumption of each tier is optimized through Lagrangian dual decomposition problem and then the hover time of UAVs is optimized through Lagrangian method. The optimality results are compared with the equally powered users in an all UHF three-tiered HetNet.

The rest of this paper is organized as follows. System model is discussed in Section II. In Section III, we present the problem formulation of optimizing the hover time of UAVs along with power of each tier. In Section IV, simulation results are presented and we conclude the paper in Section V.

\begin{figure}[!t]
\centering
\includegraphics[width=3.5in]{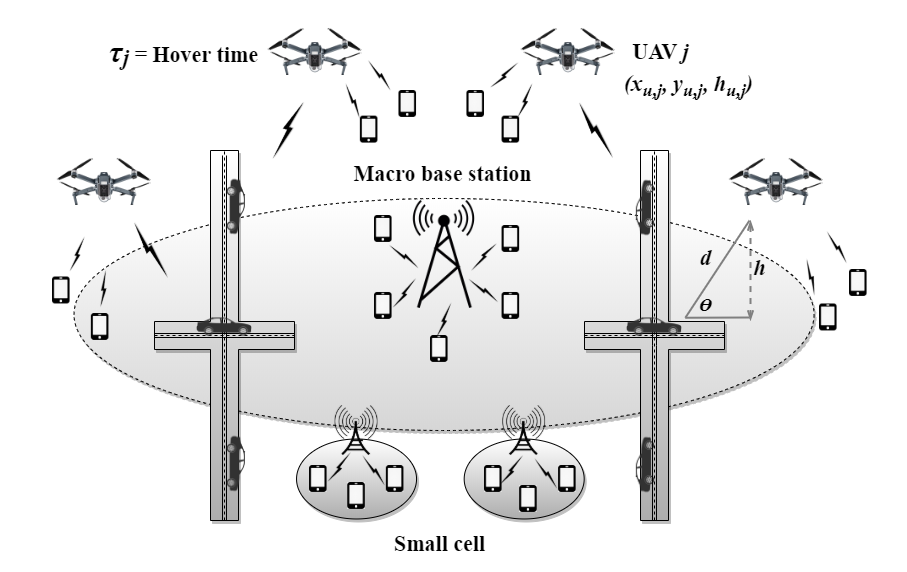}
\caption{A snapshot of system layout with a MBS overlaid with SCs and UAVs.}
\label{fig_sim}
\end{figure}

\section{System Model}
Consider the downlink transmission in a multi-tier HetNet consisting of \textit{k} tiers such that \textit{k $\in$ K = \{m,u,s\}} where the tier \textit{m} is for a MBS, the tier \textit{u} for UAVs and \textit{s} for SCs placed randomly in a geographical area having \textit{N} users (ground cellular users or vehicles), as shown in Fig. 1. There are a total of \textit{M} base stations such that $\mathbb{M}$ = \{\textit{M\textsubscript{m}, M\textsubscript{u}, M\textsubscript{s}}\}, and $|\mathbb{M}|$ = \textit{M}, where $|\cdot|$ denotes the cardinality operator. All three tiers are operating on ultra-high frequency (UHF) band where each tier \textit{k} shares the same bandwidth \textit{B} with the other two tiers. \color{black} The bandwidth \textit{B} is divided into $L$ subcarriers such that $\ell \in \{1,2,...,L\}$. Each tier assigns an exclusive subcarrier to each user within a tier \textit{k}, such that no co-tier interference exists but due to sharing of bandwidth among all tiers, the cross-tier interference is present. The subcarriers are assigned in a manner that the cross-tier interference is minimized. Each tier tries to assign such a subcarrier to its user which faces the least of interference. \color{black} The user coordinates are given by (\textit{x\textsubscript{n}, y\textsubscript{n}}) where \textit{n $\in$} \{1,2,...,\textit{N}\}. The location of MBS is denoted by (\textit{x\textsubscript{m}, y\textsubscript{m}}), while the locations of SCs are denoted by (\textit{x\textsubscript{s,j}, y\textsubscript{s,j}}) where \textit{j $\in$} \{1,2,...,\textit{S}\}, and $|\textit{M\textsubscript{s}}|$ = \textit{S}. The UAVs' locations are given by (\textit{x\textsubscript{u,j}, y\textsubscript{u,j}, h\textsubscript{u,j}}), where \textit{h\textsubscript{u,j}} represents the altitude of UAV and \textit{j $\in$ }\{1,2,...,\textit{U}\}, $|\textit{M\textsubscript{u}}|$ = \textit{U}. \textcolor{black}{It is noted that we consider the quasi-stationary UAV scenario, as in \cite{r1, r2}, where the locations of the users and UAVs are static during each communication period of interest. In case of highly mobile users, UAVs can dynamically cache the popular contents, track the mobility pattern of the corresponding
users, and then effectively serve them as noted in \cite{r1}.}

The hover time of the \textit{j\textsuperscript{th}} UAV base station is represented by $\tau$\textit{\textsubscript{j}}, which represents the time that a UAV consumes while making connections and transmitting data to the users. Optimizing the hover time extends the total flight time of UAVs. Each base station \textit{j} of tier \textit{k} has a maximum transmit power of \textit{P\textsubscript{j,k}\textsuperscript{max}}, where \textit{j $\in$ M\textsubscript{k}} for \textit{k $\in$ \{m,u,s\}}.

The path loss for an \textit{n\textsuperscript{th}} user associated with MBS tier or with \textit{j\textsuperscript{th}} base station of SC tier is given by
\begin{equation}
\color{black} PL\textit{\textsubscript{n,j}\textsuperscript{k}} = \Big(\frac{4\pi f\textsubscript{c}}{c}\Big)^2 \big(d\textit{\textsubscript{n,j}\textsuperscript{k}}\big)^{\alpha\textsubscript{\textit{k}}} \psi, \color{black}
\end{equation}
where $\alpha\textsubscript{\textit{k}}$ is the path loss exponent (PLE) and \textit{d\textsubscript{n,j}\textsuperscript{k}} is the distance between \textit{n\textsuperscript{th}} user and \textit{j\textsuperscript{th}} BS for tier \textit{k $\in$ \{m,s\}}, \textit{f\textsubscript{c}} is the carrier frequency, and $\psi$ is the log normal shadowing variable.

\color{black} The path loss for an \textit{n\textsuperscript{th}} user associated with \textit{j\textsuperscript{th}} BS of UAV tier is given by \cite{r2} and \cite{r8}, \color{black}
\begin{equation}
PL\textit{\textsubscript{n,j}\textsuperscript{u}} = \kappa\textsubscript{o} (d\textit{\textsubscript{n,j}\textsuperscript{u}})^2 \big(P\textsubscript{LoS,\textit{n}} \mu\textsubscript{LoS} + P\textsubscript{NLoS,\textit{n}} \mu\textsubscript{NLoS} \big),
\end{equation}
where  $\kappa\textsubscript{o} = \Big(\frac{4\pi f\textsubscript{c}}{c}\Big)^2$, $\mu\textsubscript{LoS}$ and $\mu\textsubscript{NLoS}$ are the additional attenuation factors for LoS and NLoS connections, and \textit{d\textit{\textsubscript{n,j}\textsuperscript{u}}} is the  distance between \textit{n\textsuperscript{th}} user and \textit{j\textsuperscript{th}} BS of tier \textit{u} and is given as $d = \sqrt{\big(\textit{x\textsubscript{u,j} - x\textsubscript{n}}\big)^2 + \big(\textit{y\textsubscript{u,j} - y\textsubscript{n}}\big)^2 + \textit{h\textsubscript{u,j}}^2}$. The probability of having LoS and NLoS connections depends upon certain factors like elevation angle between user and UAV, the atmospheric effects and the position of UAV with respect to users. \color{black} The probability of LoS connection is given by \cite{r8},\color{black} 
\begin{equation}
P\textsubscript{LoS,\textit{n}} = \frac{1}{1 + a exp \big(-b[\theta-a]\big) },
\end{equation}
where \textit{a} and \textit{b} are constants to incorporate the atmospheric effects(rural, urban, or dense urban etc.). The angle of elevation $\theta$ is given by $\theta = \frac{180}{\pi} \sin^{-1} \big(\frac{\textit{h\textsubscript{u,j}}}{d\textit{\textsubscript{n,j}\textsuperscript{u}}}\big)$. UAV forms better LoS connections when the angle of elevation is close to 90\textsuperscript{o}, but as the value of $\theta$ decreases, the probability of LoS connections decreases as well. Moreover, the probability of NLoS is given as $P\textsubscript{NLoS} = 1 - P\textsubscript{LoS}$.

The achievable rate of user \textit{n} using subcarrier $\ell$ associated with \textit{j\textsuperscript{th}} BS of tier \textit{k $\in$ K = \{m,u,s\}} is given as
\begin{equation}
R\textit{\textsubscript{n,j}\textsuperscript{k}}[\ell] = B\textsubscript{$\ell$} log\textsubscript{2} \big(1 + \gamma\textit{\textsubscript{n,j}\textsuperscript{k}}[\ell] p\textit{\textsubscript{n,j}\textsuperscript{k}}[\ell]\big),
\end{equation}
where \textit{B\textsubscript{$\ell$}} is the bandwidth assigned to each subcarrier. Considering total bandwidth that each BS gets is \textit{B} and the total number of subcarriers available to any BS are \textit{L}, we have $B\textsubscript{$\ell$} = \frac{B}{L}$. \color{black} The transmit power of user \textit{n} using subacarrier $\ell$ associated with \textit{j\textsuperscript{th}} BS of tier \textit{k $\in$ \{m,u,s\}} is \textit{p\textit{\textsubscript{n,j}\textsuperscript{k}}}[\textit{$\ell$}], \color{black} and \textit{$\gamma$\textit{\textsubscript{n,j}\textsuperscript{k}}}[\textit{$\ell$}] is given as
\begin{equation}
\gamma\textit{\textsubscript{n,j}\textsuperscript{k}}[\ell] = \frac{\big|h\textit{\textsubscript{n,j}\textsuperscript{k}}[\ell]\big|^2\big/PL\textit{\textsubscript{n,j}\textsuperscript{k}}}{I\textit{\textsubscript{n,j}\textsuperscript{k}}[\ell] + N\textsubscript{o}}.
\end{equation}
In the above equation, $\big|h\textit{\textsubscript{n,j}\textsuperscript{k}}[\ell]\big|^2$ is the squared envelope of multipath fading, \textit{N\textsubscript{o}} is the thermal noise power, and PL\textit{\textsubscript{n,j}\textsuperscript{k}} is the average path loss between \textit{n\textsuperscript{th}} user associated with \textit{j\textsuperscript{th}} BS of tier \textit{k $\in$ \{m,u,s\}}. The \textit{I\textsubscript{n,j}\textsuperscript{k}}[\textit{$\ell$}] is the total cross-tier interference received on subcarrier $\ell$ of user \textit{n} associated with \textit{j\textsuperscript{th}} BS of tier \textit{k} which is being shared by a user associated with any of the other two tiers. The cross-tier interference experienced by \textit{i\textsuperscript{th}} user associated with \textit{m\textsuperscript{th}} BS on subcarrier $\ell$ is given by
\begin{equation}
I\textit{\textsubscript{i,m}\textsuperscript{k\textsubscript{o}}}[\ell] = \sum_{\substack{k\in K \\ k \neq k\textit{\textsubscript{o}}}} \sum_{\substack{j=1 \\ j \neq m}}^{M\textit{\textsubscript{k}}} \sum_{\substack{n=1 \\ n \neq i}}^{N\textit{\textsubscript{k}}} \sigma\textit{\textsubscript{n,j}\textsuperscript{k}}[\ell] p\textit{\textsubscript{n,j}\textsuperscript{k}}[\ell] \rho\textit{\textsubscript{n,j}\textsuperscript{k}}[\ell],
\end{equation}
where $\sigma\textit{\textsubscript{n,j}\textsuperscript{k}}[\ell] = 1$ if the subcarrier is being shared by any other user, otherwise $\sigma\textit{\textsubscript{n,j}\textsuperscript{k}}[\ell] = 0$, and $\rho$\textit{\textsubscript{n,j}\textsuperscript{k}}[$\ell$] is given by $\rho\textit{\textsubscript{n,j}\textsuperscript{k}}[\ell] = \frac{\big|h\textit{\textsubscript{n,j}\textsuperscript{k}}[\ell]\big|^2}{PL\textit{\textsubscript{n,j}\textsuperscript{k}}}$. Each user consumes a circuit power equaling to \textit{P\textsubscript{c}} when a connection is established between a user and its BS. The total system power consumption can be calculated as
\begin{equation}
P\textit{\textsubscript{total}} = \sum_{k\in K} \sum_{j=1}^{M\textit{\textsubscript{k}}} \sum_{n=1}^{N\textit{\textsubscript{k}}} \sum_{\ell=1}^{L\textit{\textsubscript{k}}} p\textit{\textsubscript{n,j}\textsuperscript{k}}[\ell] + (N \times P\textsubscript{c}),
\end{equation}
whereas the system EE is defined as
\begin{equation}
EE = \frac{\sum\limits_{k\in K} \sum\limits_{j=1}^{M\textit{\textsubscript{k}}} \sum\limits_{n=1}^{N\textit{\textsubscript{k}}} \sum\limits_{\ell=1}^{L\textit{\textsubscript{k}}} R\textit{\textsubscript{n,j}\textsuperscript{k}}[\ell]}{\sum\limits_{k\in K} \sum\limits_{j=1}^{M\textit{\textsubscript{k}}} \sum\limits_{n=1}^{N\textit{\textsubscript{k}}} \sum\limits_{\ell=1}^{L\textit{\textsubscript{k}}} p\textit{\textsubscript{n,j}\textsuperscript{k}}[\ell] + (N \times P\textsubscript{c})} .
\end{equation}

\section{Problem Formulation}
The proposed methodology here is to optimize the power consumption of each tier simultaneously hence forming a sub-optimal system which will lead to the optimization of hover time of UAVs. Optimizing the average hover time will enable us to provide service to more users and for a longer duration. The power of each tier is optimized so that the system energy efficiency can be increased. The EE of each tier is optimized by having a constraint on the maximum transmit power \textit{P\textsubscript{j,k}\textsuperscript{max}} of the \textit{j\textsuperscript{th}} BS, minimum rate requirement, \textit{R\textsubscript{min}}, of user \textit{n}, and a threshold on the cross-tier interference caused on user \textit{n} associated with \textit{j\textsuperscript{th}} BS of tier \textit{k}. \color{black} In contrast to the work in \cite{r5}, we have taken into consideration the cross-tier interference constraint while determining the optimal transmit power of a user \textit{n}. Considering this constraint into our optimization regime helps to enhance the energy efficiency in an even more optimal manner, so that the number of outages occurring due to interfering powers can be mitigated in a better way. The transmit power of user \textit{n} associated with \textit{j\textsuperscript{th}} BS of tier \textit{k $\in$ \{m,u,s\}} using subcarrier $\ell$ is defined as \textit{p\textit{\textsubscript{n,j}\textsuperscript{k}}}[\textit{$\ell$}]. The transmit power of every user in each tier is optimized to form an energy efficient system. \color{black}

\subsection{Power Allocation for MBS Tier}
\color{black} The objective of our work is to maximize the system EE, thus maximizing the throughput while minimizing the power consumed, which is evident from (8). \color{black} Therefore, the objective equation for solving the EE optimization problem for MBS tier is formulated as
\begin{multline}
\max\limits_{p\textit{\textsubscript{n,j}}} EE = \max\limits_{p\textit{\textsubscript{n,j}}} \bigg[\sum\limits_{j=1}^{M\textit{\textsubscript{m}}} \sum\limits_{n=1}^{N\textit{\textsubscript{m}}} R\textit{\textsubscript{n,j}\textsuperscript{m}}[\ell] - \Big( \sum\limits_{j=1}^{M\textit{\textsubscript{m}}} \sum\limits_{n=1}^{N\textit{\textsubscript{m}}} p\textit{\textsubscript{n,j}\textsuperscript{m}}[\ell] + \\  (N \times P\textsubscript{c})\Big)\bigg], \quad \forall \ell
\end{multline}
under the constraints
\begin{equation}
\begin{gathered}
C1: \ \sum\limits_{j=1}^{M\textit{\textsubscript{m}}} \sum\limits_{n=1}^{N\textit{\textsubscript{m}}} p\textit{\textsubscript{n,j}\textsuperscript{m}}[\ell]  \leq \textit{P\textsubscript{max}\textsuperscript{j}}, \quad \forall \ell \hfill \\
\color{black} C2: \ R\textit{\textsubscript{n,j}\textsuperscript{m}}[\ell]  \geq R\textsubscript{min}, \quad \forall n,j,\ell \color{black} \hfill \\
C3: \ \sum_{\substack{k\in K \\ k \neq m}} \sum_{j=1}^{M\textit{\textsubscript{k}}} \sum_{n=1}^{N\textit{\textsubscript{k}}} \sigma\textit{\textsubscript{n,j}\textsuperscript{k}}[\ell] p\textit{\textsubscript{n,j}\textsuperscript{k}}[\ell] \rho\textit{\textsubscript{n,j}\textsuperscript{k}}[\ell]  \leq I\textsubscript{\textit{th}}[\ell], \quad \forall \ell
\end{gathered}
\end{equation}
where C1 limits the total transmit power of users associated with \textit{j\textsuperscript{th}} BS, C2 ensures the minimum rate requirement of user \textit{n}, and C3 limits the interference experienced by the user.

We solve the above optimization problem through the Lagrangian function which is given as
\begin{multline}
\mathcal{L}(p,\boldsymbol{\mu},\boldsymbol{\lambda},\boldsymbol{\phi}) = \sum\limits_{j=1}^{M\textit{\textsubscript{m}}} \sum\limits_{n=1}^{N\textit{\textsubscript{m}}} \sum\limits_{\ell=1}^{L\textit{\textsubscript{m}}}
B\textsubscript{$\ell$} log\textsubscript{2} \big(1 + \gamma\textit{\textsubscript{n,j}\textsuperscript{m}}[\ell] p\textit{\textsubscript{n,j}\textsuperscript{m}}[\ell]\big) - \\ \Big( \sum\limits_{j=1}^{M\textit{\textsubscript{m}}} \sum\limits_{n=1}^{N\textit{\textsubscript{m}}} \sum\limits_{\ell=1}^{L\textit{\textsubscript{m}}} p\textit{\textsubscript{n,j}\textsuperscript{m}}[\ell] + (N \times P\textsubscript{c})\Big) + \sum\limits_{j=1}^{M\textit{\textsubscript{m}}} \sum\limits_{n=1}^{N\textit{\textsubscript{m}}} \mu\textit{\textsubscript{n,j}} \Big(\sum\limits_{\ell=1}^{L\textit{\textsubscript{m}}} B\textsubscript{$\ell$} log\textsubscript{2} \\ \big(1 + \gamma\textit{\textsubscript{n,j}\textsuperscript{m}}[\ell] p\textit{\textsubscript{n,j}\textsuperscript{m}}[\ell]\big) - R\textsubscript{min}\Big) + \sum\limits_{j=1}^{M\textit{\textsubscript{m}}} \lambda\textsubscript{\textit{j}} \Big(\textit{P\textsubscript{max}\textsuperscript{j}} - \sum\limits_{n=1}^{N\textit{\textsubscript{m}}} \sum\limits_{\ell=1}^{L\textit{\textsubscript{m}}} p\textit{\textsubscript{n,j}\textsuperscript{m}}[\ell]\Big) + \\ \sum\limits_{\ell=1}^{L\textit{\textsubscript{k}}} \phi\textsubscript{$\ell$} \Big(I\textsubscript{\textit{th}}[\ell] - \sum_{\substack{k\in K \\ k \neq m}} \sum_{j=1}^{M\textit{\textsubscript{k}}} \sum_{n=1}^{N\textit{\textsubscript{k}}} \sigma\textit{\textsubscript{n,j}\textsuperscript{k}}[\ell] p\textit{\textsubscript{n,j}\textsuperscript{k}}[\ell] \rho\textit{\textsubscript{n,j}\textsuperscript{k}}[\ell]\Big).
\end{multline}

The above Lagrangian function is solved as two sub-problems. Firstly the optimal power is calculated by applying the 
Karush-Kahn-Tucker (KKT) conditions and then solving the Lagrange multipliers through the sub-gradient method. \color{black} The KKT conditions are the first order necessary conditions for the solution of a nonlinear constrained optimization problem to be optimal. Now by applying the KKT conditions, the partial derivative of the Lagrangian function in (11) w.r.t. $p\textit{\textsubscript{n,j}\textsuperscript{m}}[\ell]$ is set to zero to obtain the optimal solution, we have
\begin{equation}
\pdv{\mathcal{L}(p,\boldsymbol{\mu},\boldsymbol{\lambda},\boldsymbol{\phi})}{p\textit{\textsubscript{n,j}\textsuperscript{m}}[\ell]} \bigg|_{p\textit{\textsubscript{n,j}\textsuperscript{m}}[\ell] = p\textit{\textsubscript{n,j}\textsuperscript{m}}[\ell]^{*}} = 0,
\end{equation}
such that $p\textit{\textsubscript{n,j}\textsuperscript{m}}[\ell]^{*}$ is the optimal transmit power for user \textit{n} of MBS tier which is given as
\begin{equation}
p\textit{\textsubscript{n,j}\textsuperscript{m}}[\ell]^{*} = \bigg[\frac{(1+\mu\textit{\textsubscript{n,j}})B\textsubscript{$\ell$}}{\big[(1+\lambda\textsubscript{\textit{j}})+\phi\textsubscript{$\ell$}\rho\textit{\textsubscript{n,j}\textsuperscript{k}}[\ell]\big]\ln2} - \frac{1}{\gamma\textit{\textsubscript{n,j}\textsuperscript{m}}[\ell]}\bigg]^{+},
\end{equation}
where $[a]^{+} = \max(0,a)$. \color{black} The Lagrange multipliers $\mu$, $\lambda$, and $\phi$ can be computed using sub-gradient method and are given as follows
\begin{multline}
\mu\textit{\textsubscript{n,j}}(i+1) = \bigg[\mu\textit{\textsubscript{n,j}}(i) - c\textsubscript{1} \big(R\textit{\textsubscript{n,j}\textsuperscript{m}}[\ell] - R\textsubscript{min}\big)\bigg]^{+}, \qquad \\ \lambda\textsubscript{\textit{j}}(i+1) = \bigg[\lambda\textsubscript{\textit{j}}(i) - c\textsubscript{2} \big(\textit{P\textsubscript{max}\textsuperscript{j}} - \sum\limits_{n=1}^{N\textit{\textsubscript{m}}} \sum\limits_{\ell=1}^{L\textit{\textsubscript{m}}} p\textit{\textsubscript{n,j}\textsuperscript{m}}[\ell]\big)\bigg]^{+}, \hfill \\  \phi\textsubscript{$\ell$}(i+1) = \bigg[ \phi\textsubscript{$\ell$}(i) - c\textsubscript{3} \big(I\textsubscript{\textit{th}}[\ell] - \sum_{\substack{k\in K \\ k \neq m}} \sum_{j=1}^{M\textit{\textsubscript{k}}} \sum_{n=1}^{N\textit{\textsubscript{k}}} \sigma\textit{\textsubscript{n,j}\textsuperscript{k}}[\ell] p\textit{\textsubscript{n,j}\textsuperscript{k}}[\ell] \hfill \\ \rho\textit{\textsubscript{n,j}\textsuperscript{k}}[\ell]\big)\bigg]^{+},
\end{multline}
where c\textsubscript{1}, c\textsubscript{2}, and c\textsubscript{3} are the step sizes for updating Lagrange multipliers until they converge and \textit{i} denotes the iteration number. 
\textcolor{black}{It is noted that  $\mu\textit{\textsubscript{n,j}}$, $\lambda\textsubscript{\textit{j}}$ and $\phi\textsubscript{$\ell$}$ are always non-negative by the $[\cdot]^+$ function.}

\color{black} The proposed algorithm for MBS power allocation scheme is given in Algorithm 1 and it works in the following way. We take the Lagrange multipliers $\mu$\textit{\textsubscript{n,j}}, $\lambda$\textsubscript{\textit{j}}, $\phi$\textsubscript{$\ell$} as the inputs to the algorithm. And $\rho$\textit{\textsubscript{n,j}\textsuperscript{k}}[$\ell$] which is given as $\rho\textit{\textsubscript{n,j}\textsuperscript{k}}[\ell] = \frac{\big|h\textit{\textsubscript{n,j}\textsuperscript{k}}[\ell]\big|^2}{PL\textit{\textsubscript{n,j}\textsuperscript{k}}}$ and $\gamma$\textit{\textsubscript{n,j}\textsuperscript{k}}[$\ell$] as given in (5) are also taken as inputs to the algorithm. The output of the algorithm is $p\textit{\textsubscript{n,j}\textsuperscript{m}}[\ell]^{*}$ which is the optimal transmit power for user \textit{n} of MBS tier. Firstly we initialize the Lagrange multipliers and the step sizes for the iterative computation of Lagrange multipliers. Here \textit{i} is the iteration number, with \textit{i\textsubscript{max}} being the maximum number of iterations. In step 1, the algorithm will continue to run until the Lagrange multipliers have not converged or till the maximum number of iterations. We have set the convergence threshold $\epsilon$ to 0.01, so when the difference between two consecutive values of multipliers is less than $\epsilon$, then we will obtain the optimal transmit power for user \textit{n}. At the start of the algorithm, step 2 will update the $p\textit{\textsubscript{n,j}\textsuperscript{m}}[\ell]^{*}$ based on the initial set values and then in step 3 the $\mu$, $\lambda$ and $\phi$ are updated based on the value computed in step 2. For a constant step size, our algorithm converges to a value within some range of optimal value, and it may not necessarily be the optimal value, i.e. \textit{p\textsubscript{optimal} - p\textsubscript{obtained} $<$ $\epsilon$} \cite{r28}. \color{black} 

\color{black}
\begin{algorithm}[t]
	\color{black}
	\caption{Power allocation for MBS tier}
	\renewcommand{\algorithmicrequire}{\textbf{Input:}}
	\renewcommand{\algorithmicensure}{\textbf{Output:}}
	\begin{algorithmic}[1]
		\color{black}	
		\REQUIRE $\mu\textit{\textsubscript{n,j}}, \lambda\textsubscript{\textit{j}}, \phi\textsubscript{$\ell$}, B\textsubscript{$\ell$}, \rho\textit{\textsubscript{n,j}\textsuperscript{k}}[\ell], \gamma\textit{\textsubscript{n,j}\textsuperscript{m}}[\ell]$.
		\ENSURE  $p\textit{\textsubscript{n,j}\textsuperscript{m}}[\ell]$.
		
		Set $\mu\textit{\textsubscript{n,j}} = \lambda\textsubscript{\textit{j}} = \phi\textsubscript{$\ell$} = 0.01, \enspace c\textsubscript{1} = c\textsubscript{2} = c\textsubscript{3} = 0.01, \enspace i = 1, \enspace i\textsubscript{max} = 10^{6}, \epsilon = 0.01 $
		
		\WHILE {$\mu\textit{\textsubscript{n,j}}$, $\lambda\textsubscript{\textit{j}}$, and $\phi\textsubscript{$\ell$}$ have not converged or $i\leq i\textsubscript{max}$}
		\STATE Calculate $p\textit{\textsubscript{n,j}\textsuperscript{k}}[\ell]$ from (13).
		\STATE Update $\mu\textit{\textsubscript{n,j}}$, $\lambda\textsubscript{\textit{j}}$, and $\phi\textsubscript{$\ell$}$ from (14).
		\ENDWHILE
		
		\hspace{-1em}\textbf{End}
		\color{black}
	\end{algorithmic}
\color{black}
\end{algorithm}
\color{black}

\subsection{Power Allocation for UAV and SC Tier}
The power allocation for UAV and SC tiers is optimized in a similar way as the MBS tier. The optimal transmit power for UAV and SC tier is given as
\begin{equation}
p\textit{\textsubscript{n,j}\textsuperscript{k}}[\ell]^{*} = \bigg[\frac{(1+\mu\textit{\textsubscript{n,j}})B\textsubscript{$\ell$}}{\big[(1+\lambda\textsubscript{\textit{j}})+\phi\textsubscript{$\ell$}\rho\textit{\textsubscript{n,j}\textsuperscript{k}}[\ell]\big]\ln2} - \frac{1}{\gamma\textit{\textsubscript{n,j}\textsuperscript{k}}[\ell]}\bigg]^{+},
\end{equation}
where \textit{k$\in$\{u,s\}}.

\subsection{UAV Hover Time Optimization}
A UAV has finite source of energy, which elevates the need to optimize its battery usage so that it can service the users for a longer duration. Hence it becomes vital to optimize the hover time of UAVs. Hover time of a UAV is the time it takes to associate with the users through control signaling and then transmitting data. The hover time of a \textit{j\textsuperscript{th}} UAV is represented by $\tau$\textit{\textsubscript{j}} and we will represent the number of users associated with a \textit{j\textsuperscript{th}} UAV as \textit{N\textsubscript{j}} utilizing \textit{L\textsubscript{j}} subcarriers, where
\begin{equation}
\tau\textit{\textsubscript{j}} = \sum_{n=1}^{N\textit{\textsubscript{j}}} \sum\limits_{\ell=1}^{L\textit{\textsubscript{j}}} T\textit{\textsubscript{n,j}\textsuperscript{u}}[\ell] + N\textit{\textsubscript{j}}  t\textit{\textsubscript{c}}.
\end{equation}
\color{black} In the above equation, \textit{t\textsubscript{c}} is the control signaling time of user \textit{n} and is a constant value for all users. \color{black} However, the total control time of a UAV depends upon the number of users \textit{N\textsubscript{j}} associated with \textit{j\textsuperscript{th}} UAV. \textit{T\textsubscript{n,j}\textsuperscript{u}}[$\ell$] is the data transmission time of user \textit{n} associated with \textit{j\textsuperscript{th}} UAV. The data transmission time of user \textit{n} is related to the rate as
\begin{equation}
R\textit{\textsubscript{n,j}\textsuperscript{u}}[\ell] = \frac{\beta\textit{\textsubscript{n,j}\textsuperscript{u}}[\ell]}{T\textit{\textsubscript{n,j}\textsuperscript{u}}[\ell]},
\end{equation}
where $\beta$\textit{\textsubscript{n,j}\textsuperscript{u}}[$\ell$] is the data requirement of user \textit{n} in bits. The achievable rate of user \textit{n} using subcarrier $\ell$ and associated with \textit{j\textsuperscript{th}} BS of UAV tier is
\begin{equation}
R\textit{\textsubscript{n,j}\textsuperscript{u}}[\ell] = B\textsubscript{$\ell$} log\textsubscript{2} \big(1 + \gamma\textit{\textsubscript{n,j}\textsuperscript{u}}[\ell] p\textit{\textsubscript{n,j}\textsuperscript{u}}[\ell]\big).
\end{equation}
From equation (17) and (18) we get
\begin{equation}
p\textit{\textsubscript{n,j}\textsuperscript{u}}[\ell] = \frac{2^{\frac{\beta\textit{\textsubscript{n,j}\textsuperscript{u}}[\ell]}{B\textsubscript{$\ell$} T\textit{\textsubscript{n,j}\textsuperscript{u}}[\ell]}} - 1}{\gamma\textit{\textsubscript{n,j}\textsuperscript{u}}[\ell]}.
\end{equation}

The objective equation for solving hover time optimization problem is formulated as
\begin{equation}
\max\limits_{T\textit{\textsubscript{n,j}}} \bigg[\sum\limits_{j=1}^{M\textit{\textsubscript{u}}} \sum\limits_{n=1}^{N\textit{\textsubscript{u}}} \frac{\beta\textit{\textsubscript{n,j}\textsuperscript{u}}[\ell]}{T\textit{\textsubscript{n,j}\textsuperscript{u}}[\ell]} - \sum\limits_{j=1}^{M\textit{\textsubscript{u}}} \sum\limits_{n=1}^{N\textit{\textsubscript{u}}} \bigg(\frac{2^{\frac{\beta\textit{\textsubscript{n,j}\textsuperscript{u}}[\ell]}{B\textsubscript{$\ell$} T\textit{\textsubscript{n,j}\textsuperscript{u}}[\ell]}} - 1}{\gamma\textit{\textsubscript{n,j}\textsuperscript{u}}[\ell]}\bigg) \bigg], \forall \ell
\end{equation}
under the constraints
\begin{equation}
\begin{gathered}
C1: \ \beta\textit{\textsubscript{n,j}\textsuperscript{u}}[\ell]{T\textit{\textsubscript{n,j}\textsuperscript{u}}[\ell]} \geq T\textsubscript{min}, \quad \forall n,j,\ell \\ 
C2: \ \sum\limits_{n=1}^{N\textit{\textsubscript{j}}} \sum\limits_{\ell=1}^{L\textit{\textsubscript{j}}} T\textit{\textsubscript{n,j}\textsuperscript{u}}[\ell] \leq \tau\textit{\textsubscript{j}} - T\textit{\textsubscript{c,j}}, \quad \forall j
\end{gathered}
\end{equation}
where C1 determines transmission time based on the minimum rate and load of user, while C2 ensures that the total data transmission time does not exceed maximum hover time. 

The Lagrangian function of the above optimization problem is given as
\begin{multline}
\mathcal{L}(T, \boldsymbol{\mu}, \boldsymbol{\lambda}) = \sum\limits_{j=1}^{M\textit{\textsubscript{u}}} \sum\limits_{n=1}^{N\textit{\textsubscript{u}}} \sum\limits_{\ell=1}^{L\textit{\textsubscript{u}}} \frac{\beta\textit{\textsubscript{n,j}\textsuperscript{u}}[\ell]}{T\textit{\textsubscript{n,j}\textsuperscript{u}}[\ell]} - \sum\limits_{j=1}^{M\textit{\textsubscript{u}}} \sum\limits_{n=1}^{N\textit{\textsubscript{u}}} \sum\limits_{\ell=1}^{L\textit{\textsubscript{u}}} \\  \bigg(\frac{2^{\frac{\beta\textit{\textsubscript{n,j}\textsuperscript{u}}[\ell]}{B\textsubscript{$\ell$} T\textit{\textsubscript{n,j}\textsuperscript{u}}[\ell]}} - 1}{\gamma\textit{\textsubscript{n,j}\textsuperscript{u}}[\ell]}\bigg) + \sum\limits_{j=1}^{M\textit{\textsubscript{u}}} \sum\limits_{n=1}^{N\textit{\textsubscript{u}}} \mu\textit{\textsubscript{n,j}} \bigg(\frac{\beta\textit{\textsubscript{n,j}\textsuperscript{u}}[\ell]}{T\textit{\textsubscript{n,j}\textsuperscript{u}}[\ell]} - T\textsubscript{min}\bigg)  + \sum\limits_{j=1}^{M\textit{\textsubscript{u}}} \lambda\textsubscript{\textit{j}} \\ \Big( \big(\tau\textit{\textsubscript{j}} - T\textit{\textsubscript{c,j}}\big) - \sum\limits_{n=1}^{N\textit{\textsubscript{u}}} \sum\limits_{\ell=1}^{L\textit{\textsubscript{u}}} T\textit{\textsubscript{n,j}\textsuperscript{u}}[\ell] \Big).
\end{multline}

This kind of dual-optimization problem is solved by first applying the KKT conditions to the above Lagrangian function and then utilizing the sub-gradient method to obtain the Lagrange multipliers, i.e.,
\begin{equation}
\pdv{\mathcal{L}(T,\boldsymbol{\mu},\boldsymbol{\lambda})}{T\textit{\textsubscript{n,j}}} = 0.
\end{equation}
Simplifying (23) will give us,
\begin{equation}
\frac{\beta\textit{\textsubscript{n,j}\textsuperscript{u}}[\ell]}{\big(T\textit{\textsubscript{n,j}\textsuperscript{u}}[\ell]\big)^{2}} \bigg[\frac{\ln(2) . 2^{\frac{\beta\textit{\textsubscript{n,j}\textsuperscript{u}}[\ell]}{B\textsubscript{$\ell$} T\textit{\textsubscript{n,j}\textsuperscript{u}}[\ell]}}}{B\textsubscript{$\ell$} \gamma\textit{\textsubscript{n,j}\textsuperscript{u}}[\ell]} - \big(1 + \mu\textit{\textsubscript{n,j}}\big)\bigg] - \lambda\textsubscript{\textit{j}} = 0
\end{equation}
Equation (24) is a nonlinear equation and it has no direct solution, hence we solve the equation through numerical methods to obtain the optimal hover time for UAVs. The Lagrange multipliers $\mu$ and $\lambda$ are updated using the sub-gradient method as given below
\begin{multline}
\mu\textit{\textsubscript{n,j}}(i+1) = \bigg[\mu\textit{\textsubscript{n,j}}(i) - c\textsubscript{1} \bigg( \frac{\beta\textit{\textsubscript{n,j}\textsuperscript{u}}[\ell]}{T\textit{\textsubscript{n,j}\textsuperscript{u}}[\ell]} - T\textsubscript{min}\bigg)\bigg]^{+}, \\ \lambda\textsubscript{\textit{j}}(i+1) = \bigg[\lambda\textsubscript{\textit{j}}(i) - c\textsubscript{2} \Big( \big(\tau\textit{\textsubscript{j}} - T\textit{\textsubscript{c,j}}\big) - \sum\limits_{n=1}^{N\textit{\textsubscript{u}}} \sum\limits_{\ell=1}^{L\textit{\textsubscript{u}}} T\textit{\textsubscript{n,j}\textsuperscript{u}}[\ell] \Big) \bigg]^{+},
\end{multline}
where c\textsubscript{1} and c\textsubscript{2} are step size and \textit{i} is the iteration number. \textcolor{black}{It is noted that  $\mu\textit{\textsubscript{n,j}}$ and $\lambda\textsubscript{\textit{j}}$ are always non-negative by the $[\cdot]^+$ function.}

\section{Simulation Results and Analysis}
\color{black} In our simulations, we consider a three-tier HetNet in a 1000m $\times$ 1000m rectangular region of interest (RoI) where \textit{N} = 100 users (ground cellular users or vehicles) are distributed according to a uniform distribution. \color{black}  \color{black} The MBS is located at the center of RoI, whereas UAVs and SCs are deployed at random locations, with each UAV having a fixed altitude of 140m \cite{r5,r26}. The number of SCs deployed are fixed at \textit{s = 3}, while the number of UAVs \textit{u} vary for the purpose of simulations. \color{black} Each tier shares the same bandwidth \textit{B} = 20 MHz with the other tiers, which is divided into 64 subcarriers. The MBS has a maximum transmit power of 45 dBm, while for UAVs and SCs it is 30 and 27 dBm, respectively. The minimum rate requirement of a user, \textit{R\textsubscript{min}}, is 0.25 Mbps and the interference threshold is set to be 10\textsuperscript{-14} W which is same for all users. \color{black} Each user has a load requirement of 10 Mb and the maximum hover time of a UAV is 30 minutes. The PLEs for MBS and SCs are kept different due to the topographical differences, as the indoor SC users experience a higher attenuation as compared to outdoor MBS users. Having different PLEs for different links based on the physical environment improves the performance of the system \cite{r29}. Following  \cite{r3,r26}, and \cite{r27}, we assume the path loss model parameters in Table I. \color{black}

\begin{table}[!t]
	\renewcommand{\arraystretch}{1.5}
	\caption{Simulation Parameters}
	\label{tab:Simulation_Parameters}
	\centering
	\begin{tabular}{|M{1.2cm}|M{4.8cm}|M{1.5cm}|}
		\hline
		\textbf{Parameter} & \textbf{Description} & \textbf{Value} \\
		\hline \hline
		\textit{f\textsubscript{c}}             & Carrier frequency & 2.4 GHz \\ \hline
		\textit{N\textsubscript{o}}             & Thermal noise power & -174 dBm/Hz \\ \hline
		\color{black} $\alpha$     & \color{black} Path loss exponent for MBS (dense urban, high rise urban) & \color{black} (2.5, 2.7) \\ \hline
		\color{black} $\beta$   & \color{black} Path loss exponent for SC (dense urban, high rise urban) & \color{black} (2.6, 2.9) \\ \hline
		\color{black} $\psi$             & \color{black} Log normal shadowing (dense urban, high rise urban) & \color{black} (4 dB, 6 dB) \\ \hline
		\textit{$\mu$\textsubscript{LoS}}             & Additional path loss by LoS connection (dense urban, high rise urban) & (1.6, 2.3) \\ \hline
		\textit{$\mu$\textsubscript{NLoS}}            & Additional path loss by NLoS connection (dense urban, high rise urban) & (23, 34) \\ \hline
		\textit{a,b} & Dense urban environment constants & 12.08, 0.11 \\
		\hline
		\textit{a,b} & High rise urban environment constants & 27.23, 0.08 \\
		\hline
	\end{tabular}
\end{table}

We  first analyze the effects on the system EE of the proposed three-tier HetNet under optimal and maximal power allocation schemes for increasing number of UAVs. Furthermore we study how the average hover time of UAVs is affected with optimal and equal time allocation when the number of UAVs in the network is varied. \color{black} Simulation results were averaged over $10^3$ Monte Carlo iterations performed in MATLAB. \color{black} In Fig. 2, we  consider two environment scenarios for  analysis; one is the dense urban and the other is high rise urban. In each scenario, we can observe that optimal power allocation scheme outperforms the maximal power allocation. The total system power consumption is lower in case of optimal power allocation, which reduces the cross-tier interference and improves the SINR, thus enhancing the sum rate and system EE. When the number of UAVs increases, the cell radius of UAVs becomes narrower with fewer users associating with each UAV. This enables the UAVs to establish better LoS connections with the associated users, thus improving the data rates of the edge users. We can observe a 38\% rise in the system EE when there are 10 UAVs deployed in the HetNet. As more UAVs are deployed in the network, the edge capacity of the macrocell enhances, resulting in lesser outages and better system EE.

\begin{figure}[!t]
	\centering
	\includegraphics[width=3.5in]{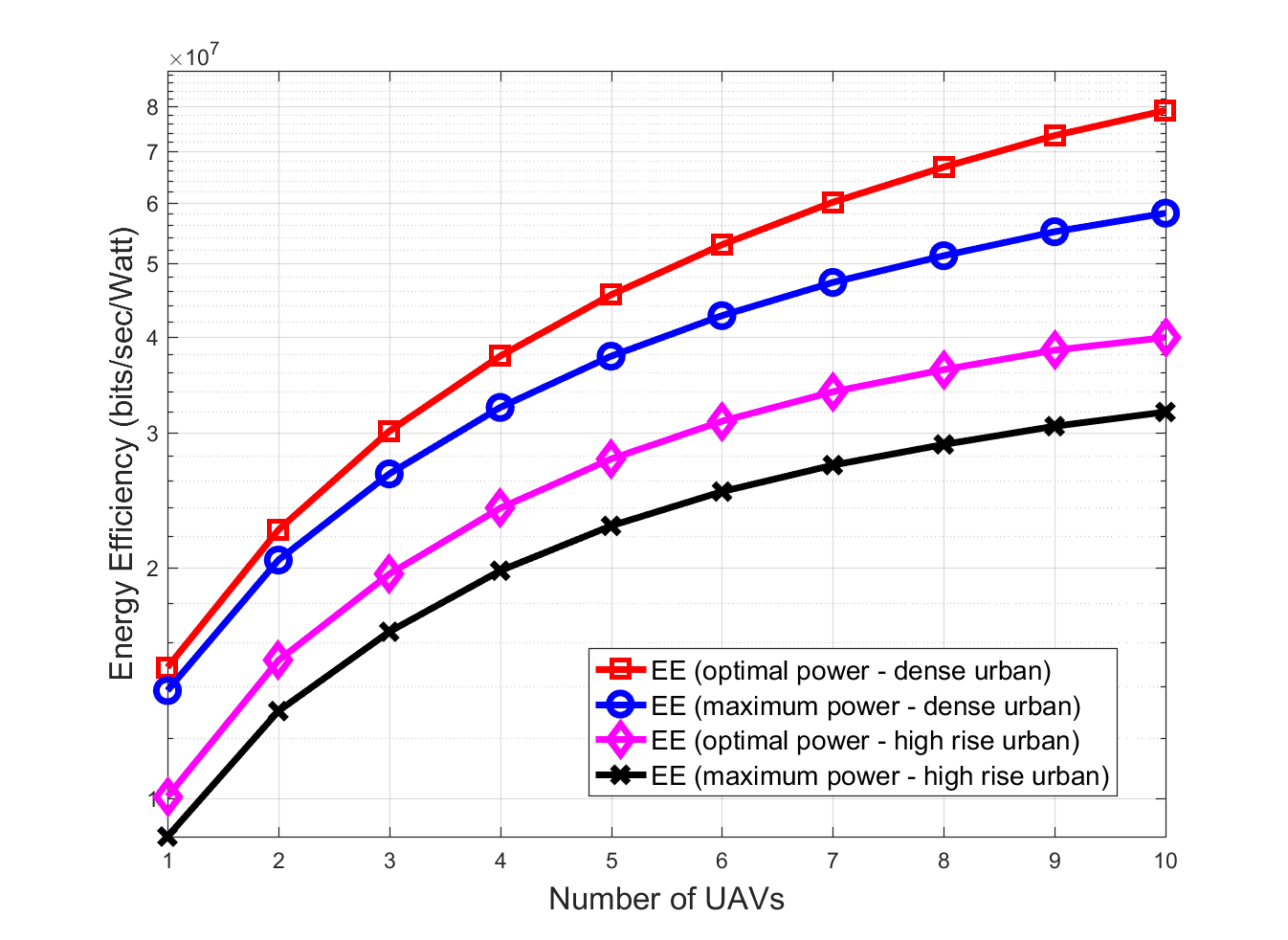}
	\caption{System EE of a HetNet for different environments with optimal and maximal power allocation with varying number of UAVs for $N=100$.}
	\label{fig_sim}
\end{figure}

Fig. 3 depicts how the system EE is affected when the user density and number of UAVs is varied. For a lower user density, the power allocated to each user is significant enough to lessen the effects of path loss attenuation and cross-tier interference, which results in a system which is efficient in energy consumption. For a higher user density and fewer UAVs in the network, the system EE decreases rapidly as the total available power is small.  With more users in the network, there is an increased cross-tier interference resulting in a degraded SINR and thereby poor data rates. To provide better coverage and enhance the system EE in case of higher user density, more UAVs need to be deployed in the network. In hotspot regions such as stadiums or arenas or traffic jams, when the number of ground cellular users or connected vehicles go beyond the coverage capacity of MBS, more UAVs should be deployed in the network to enhance the system EE and provide coverage to the added users, as depicted in Fig. 3.

\begin{figure}[!t]
	\centering
	\includegraphics[width=3.5in]{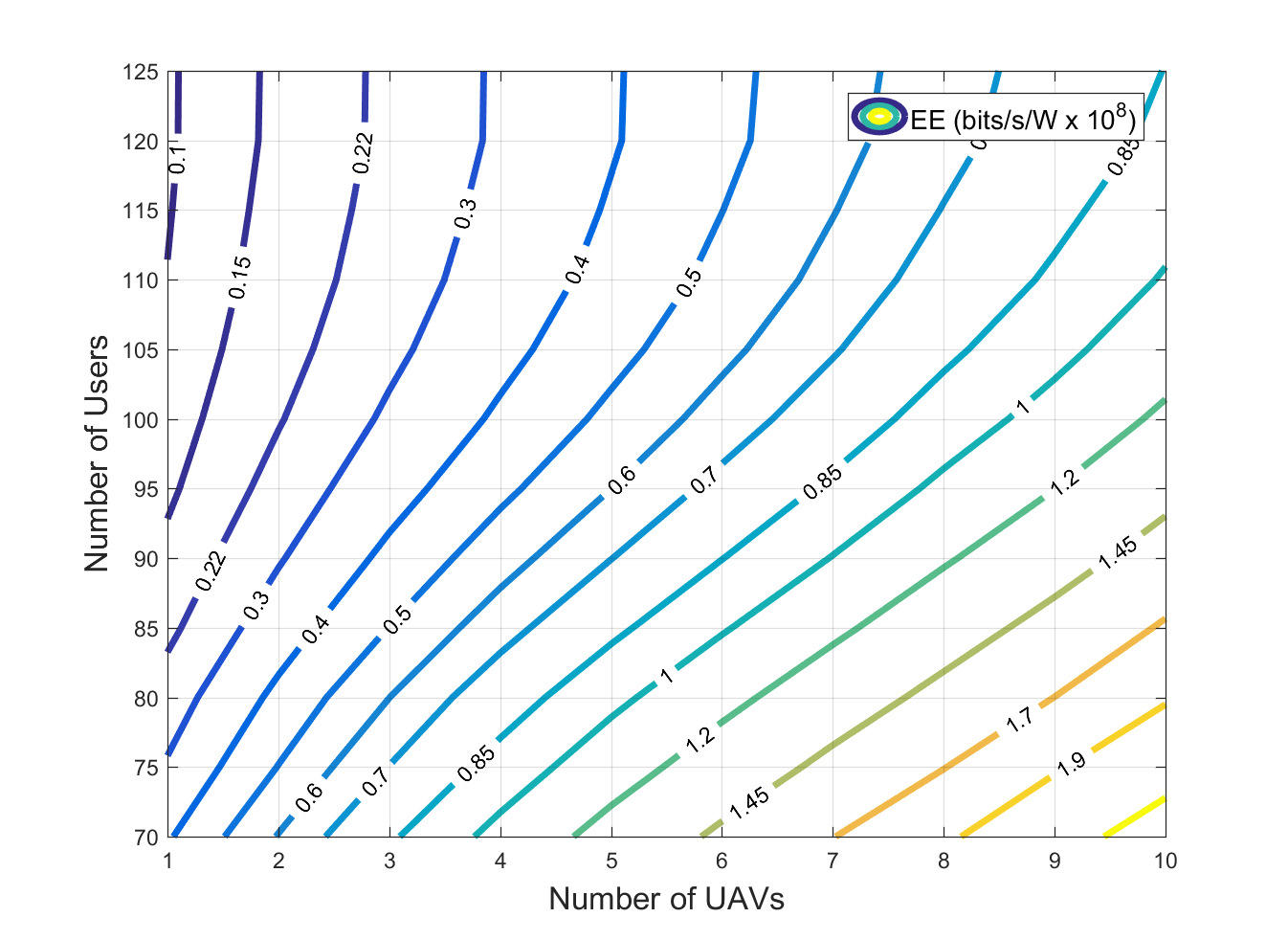}
	\caption{\color{black} System EE of a HetNet with varying number of users and UAVs with optimal power allocation in dense urban environments. \color{black} }
	\label{fig_sim}
\end{figure}

\begin{figure}[!t]
	\centering
	\includegraphics[width=3.5in]{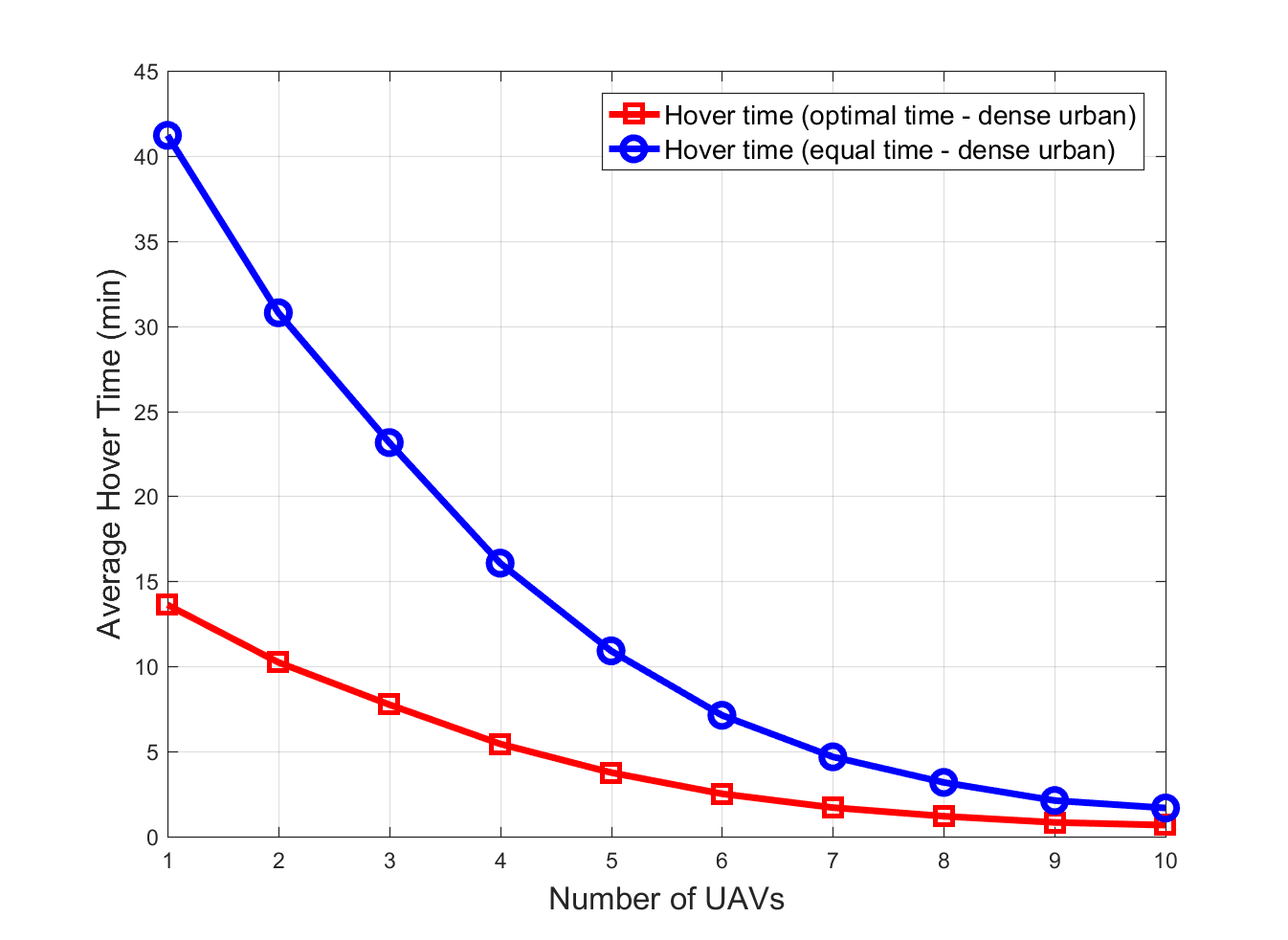}
	\caption{Average hover time of UAVs in a HetNet with optimal and equal time allocation with varying number of UAVs for $N=100$.}
	\label{fig_sim}
\end{figure}

Hover time of a UAV is the time which it takes to serve the users (ground cellular users or connected vehicles). UAVs are battery powered with intermittent source of energy, which makes it vital to utilize that energy in the most optimal manner for a longer service duration. Fig. 4 depicts the comparison of UAVs average hover time performance in a dense urban environment when optimal and equal time is allocated to users with varying number of UAVs. We can observe that with optimal energy utilization, UAVs have to hover for shorter time duration to service the users and thus making it possible to extend the total service duration. On the other hand, with the non-optimal/random time allocation, a UAV has to hover for twice as much time to service the same number of users. As more UAVs are deployed in the network, fewer users get served by each UAV and signal strength increases, causing a rise in data rates. Higher rates translate to rapid data transmission time, and a UAV has to hover for shorter duration. As the number of UAVs continues to increase, the non-optimal average hover time approaches the optimal average hover time. However, in the cases where we cannot afford to deploy a large number of UAVs,  optimal time allocation provides us a significant increase in the service duration of UAVs.

\color{black} In our simulations, each user has a fixed load requirement of 10 Mb. \color{black}  In Fig. 5, we analyze how the change in load requirement of a user affects the average hover time of UAVs when the number of UAVs is varied. When the data demands of a user increase, the data transmission time increases, and the UAV will have to hover for longer duration to serve that user. But for the same data demands, if we increase the number of UAVs, then the average hover time decreases. \color{black} For a load requirement of 10 Mb, we can observe that there is a decrease of about 64\% in the average hover time when number of UAVs is increased from one to five in the network. The energy of a single UAV servicing 20 users with load requirement of 20 Mb each will dissipate quickly as compared to two UAVs each servicing 10 users. \color{black}  When the load is shared among UAVs, then the average hover time decreases significantly, thus extending the service duration. \color{black} Furthermore, in \cite{r2} it is noted that there is a trade-off between UAV hover time and bandwidth efficiency for an increasing number of UAVs. But in our work, we have taken equal bandwidth allocation for all users, while optimizing the power consumed by each user. Optimizing power consumption of a UAV BS enhances the sum rate, which subsequently results in a longer hover duration.\color{black}

Hover time of a UAV is dependent on the number of users it connects. With each additional user, the control signaling time as well as data transmission time increases. Fig. 6 shows how the change in user density affects the average hover time of UAVs when number of UAVs are varied in the network. We can observe from the figure that as the number of users increases, the UAVs connect more users, which gives rise to control signaling time and data transmission time, causing the UAVs to hover for longer duration to service the users and their energy will dissipate quickly. If we have a disaster scenario and the UAVs are needed for the dissemination of public safety announcements and to provide coverage to the users which are disconnected from the network or during traffic jams with a sudden increase in connected vehicles,  it will be desired that UAVs provide coverage for as long as possible. In situations like this, we can adjust how many UAVs to be deployed depending on the number of users.

\begin{figure}[!t]
	\centering
	\includegraphics[width=3.5in]{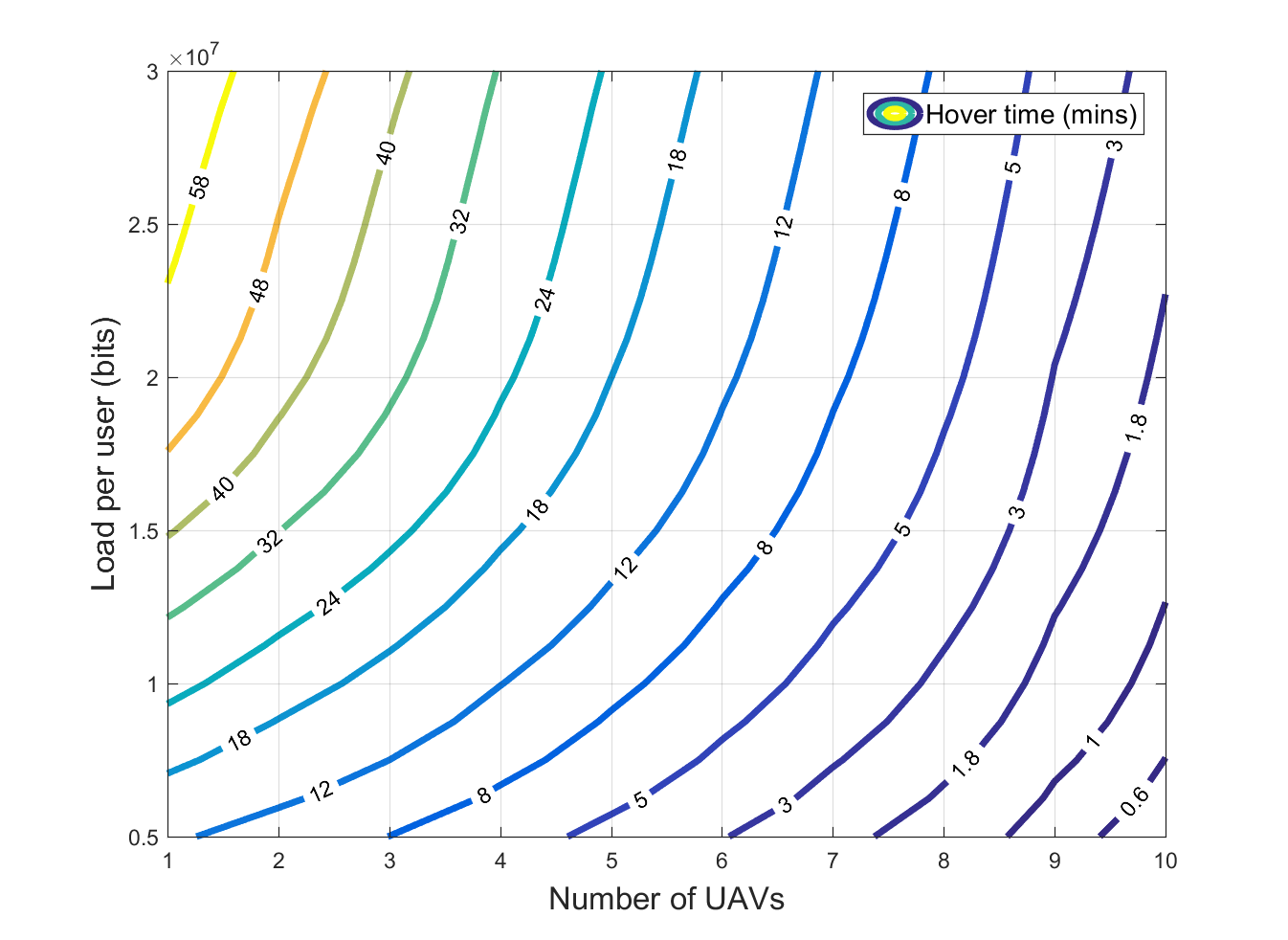}
	\caption{Average hover time of UAVs in a HetNet with varying load requirements of users and number of UAVs for $N=100$.}
	\label{fig_sim}
\end{figure}

\begin{figure}[!t]
	\centering
	\includegraphics[width=3.5in]{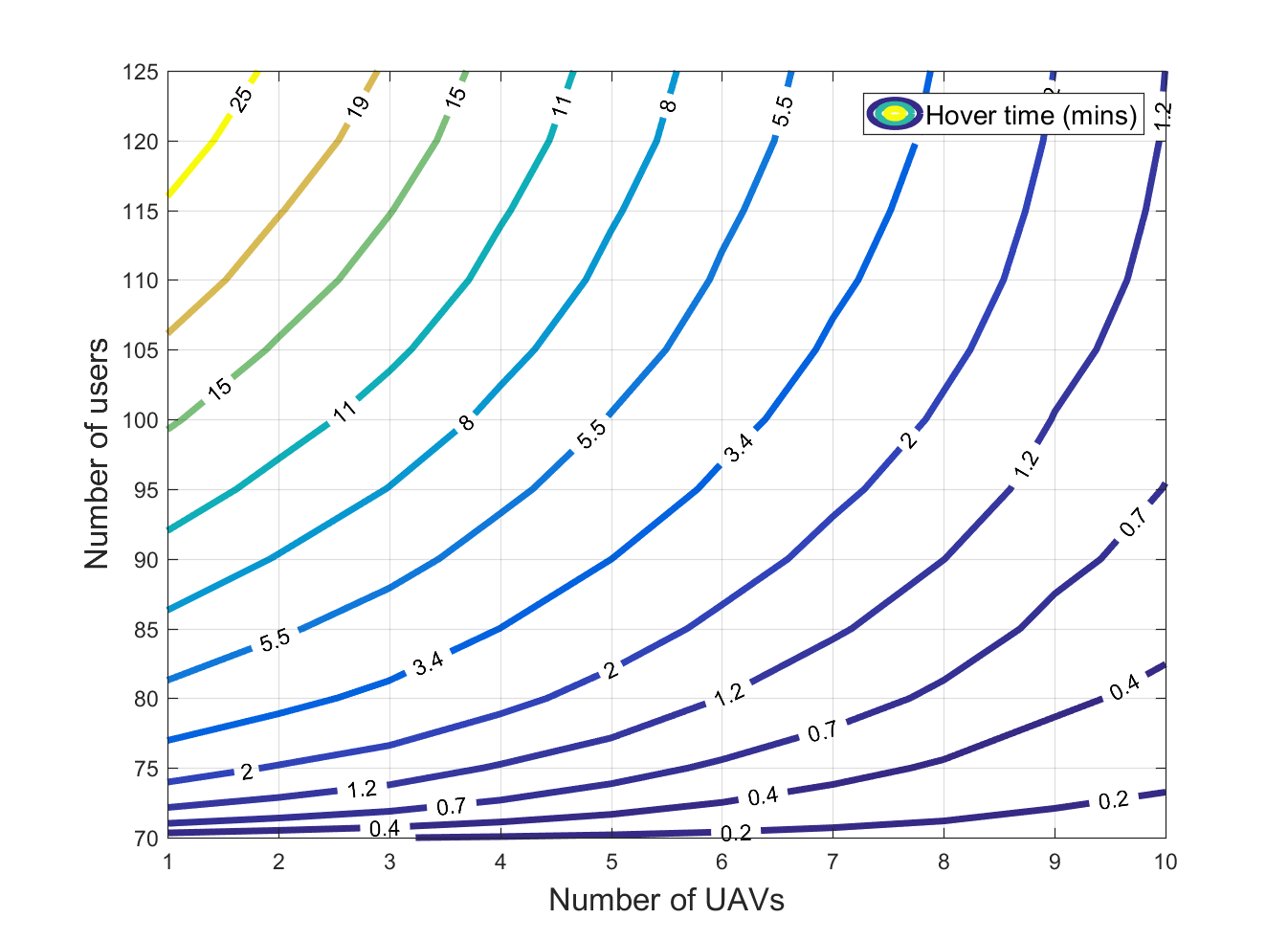}
	\caption{Average hover time of UAVs in a HetNet with varying number of users and UAVs.}
	\label{fig_sim}
\end{figure}

\begin{figure}[!t]
	\centering
	\includegraphics[width=3.5in]{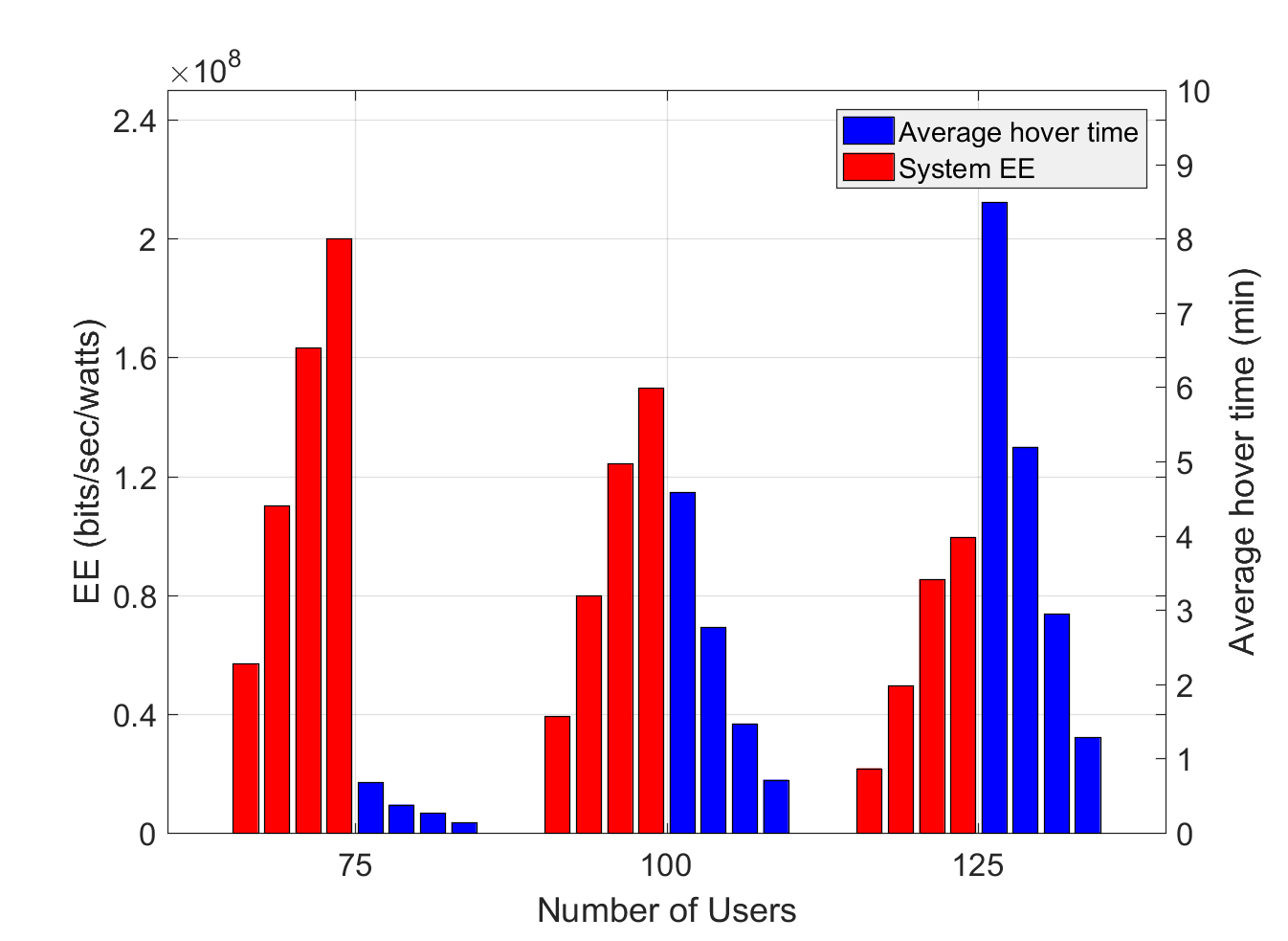}
	\caption{System EE vs average hover time comparison with varying number of users for UAVs = 3, 5, 8 and 10, respectively for each grouped bars.}
	\label{fig_sim}
\end{figure}

The motivation behind our work is scalability of network resources in particular scenarios like disaster situations, hotspot regions, and to improve the cell edge capacity of a macrocell in case of outages. Small cells are fixed devices which are employed to improve the signal quality and rates of indoor users, and they cannot be installed in real time. UAVs on the other hand are battery powered devices and can be deployed as per the network demand. Fig. 7 provides the network operator with the statistics of how the system EE and average hover time is affected by the change in user density and under different number of UAVs in the network. Our goal is to utilize the available resources to their maximum potential and conserve the energy of the system. In Fig. 7, each grouped bars represent varying number of UAVs in ascending order from 3, 5, 8 and 10, respectively.  We can observe the trend for a fixed number of UAVs that as the user density increases, the system EE starts decreasing and average hover time of UAVs is increasing. As the resources are same, therefore, the added users come at the cost of degrading the system. This graph can assist the network operators to decide how much resources to utilize for a specific user density, and whether they prefer a longer service duration and better system EE by adding more UAVs in the network or by compromising on higher data rates and having shorter flight times by deploying fewer UAVs.

\section{Conclusion}
With the increase in data demands of mobile cellular users and connected vehicles, it has become challenging to satisfy users with the available terrestrial framework. In this work, we have formulated a heterogeneous network consisting of UAVs and SCs to enhance the capacity of a macro cell and presented a two layer scheme to form an energy efficient system which consists of firstly optimizing the power consumption of each tier and then optimizing the average hover time of UAVs. This way we can conserve the energy of the system and utilize it in the most optimum way. The optimization scheme is comprised of  Lagrangian multipliers method and sub-gradient method to achieve the optimal power for each user and the optimal average hover time for UAVs. The results showed a significant enhancement in the system EE and the flight time of UAVs. This is particularly beneficial in the cases of temporary hotspot regions and emergency scenarios where we are more concerned with connectivity for longer durations rather than delivering high data rates. \textcolor{black}{The future extensions of this paper include the impacts of hardware imperfections and antenna characteristics.}


%

%

\section*{Acknowledgment}
The authors extend their appreciation to the Researchers Supporting Project number (RSP-2020/32), King Saud University, Riyadh, Saudi Arabia for funding this work.

%
%
%

\ifCLASSOPTIONcaptionsoff
  \newpage
\fi



%

\bibliographystyle{IEEEtran}

\bibliographystyle{IEEEtran}


%

\newpage
\begin{IEEEbiography}[{\includegraphics[width=1in,height=1.15in,clip,keepaspectratio]{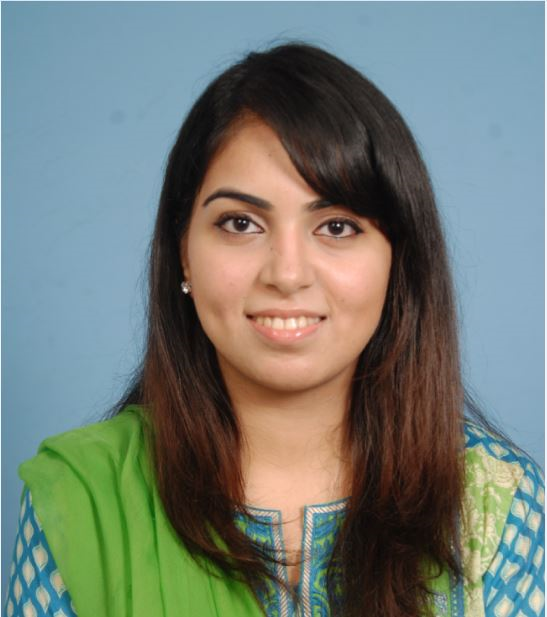}}]{Sidra Tul Muntaha} received her B.E. degree in electrical engineering (High Hons.) from National University of Computer and Emerging Sciences (NUCES) - FAST, Pakistan, in 2015 and her M.S. degree in electrical engineering (Highest Hons.) from National University of Sciences and Technology (NUST), Pakistan, in 2019. Her research interests include UAV/drone communication and 5G heterogeneous networks.        
 \end{IEEEbiography}

\vskip -2\baselineskip plus -1fil

\begin{IEEEbiography}[{\includegraphics[width=1in,height=1.25in,clip,keepaspectratio]{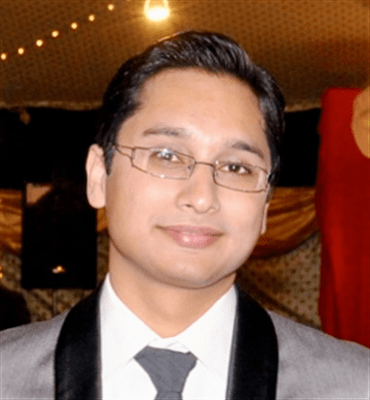}}]{Syed Ali Hassan}
(S'09-M'12-SM'17) received his PhD in Electrical Engineering from Georgia Institute of Technology, Atlanta USA in 2011. He received his MS Mathematics from Georgia Tech in 2011 and MS Electrical Engineering from University of Stuttgart, Germany in 2007. He was awarded BE degree in Electrical Engineering from National University of Sciences and Technology (NUST), Pakistan, in 2004. His broader area of research is signal processing for communications with a focus on cooperative communications for wireless networks, stochastic modeling, estimation and detection theory, and smart grid communications. Currently, he is working as an Associate Professor at the School of Electrical Engineering and Computer Science (SEECS), NUST, where he is the director of Information Processing and Transmission (IPT) Lab, which focuses on various aspects of theoretical communications. He was a visiting professor at Georgia Tech in Fall 2017 and also holds senior membership of IEEE. He also held industry positions, in Cisco Systems Inc. CA, USA, and Center for Advanced Research in Engineering, Islamabad, Pakistan.
\end{IEEEbiography}
\vskip -2\baselineskip plus -1fil

\begin{IEEEbiography}[{\includegraphics[width=1in,height=1.25in,clip,keepaspectratio]{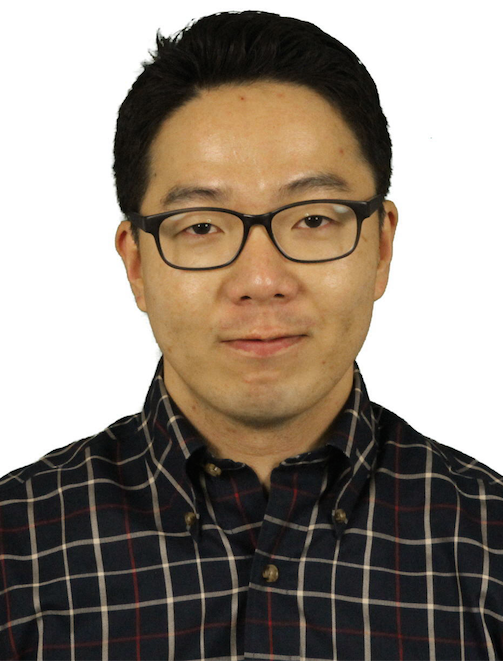}}]{Haejoon Jung}
 received the B.S. degree (Hons.) from Yonsei University, South Korea, in 2008, and the M.S. and Ph.D. degrees from the Georgia Institute of Technology (Georgia Tech), Atlanta, GA, USA, in 2010 and 2014, respectively, all in electrical engineering. From 2014 to 2016, he was a Wireless Systems Engineer at Apple, Cupertino, CA, USA. He joined Incheon National University, Incheon, South Korea, in 2016, where he is currently an Associate Professor with the Department of Information and Telecommunication Engineering. His research interests include communication theory, wireless communications, wireless power transfer, and statistical signal processing.
\end{IEEEbiography}

\begin{IEEEbiography}{M. Shamim Hossain}
received the Ph.D. degree in electrical and computer engineering from the University of Ottawa, Ottawa, ON, Canada. He is with the College of Computer and Information Sciences, King Saud University, Riyadh, Saudi Arabia. He is also an Adjunct Professor with the School of Electrical Engineering and Computer Science, University of Ottawa. His research interests include cloud networking, smart environment (smart city, smart health), AI, deep learning, edge computing, Internet of Things (IoT), multimedia for health care, and multimedia big data. He has authored and coauthored approximately 240 publications including refereed journals, conference papers, books, and book chapters. He was the Co-Chair, General Chair, Workshop Chair, Publication Chair, and TPC for over 12+ IEEE and ACM conferences and workshops. He is currently the Co-Chair of the 3rd IEEE ICME workshop on Multimedia Services and Tools for smart-health (MUST-SH 2020). He is a recipient of a number of awards, including the Best Conference Paper Award and the 2016 ACM Transactions on Multimedia Computing, Communications and Applications (TOMM) Nicolas D. Georganas Best Paper Award, and the the 2019 King Saud University Scientific Excellence Award (Research Quality). He is on the Editorial Board of the IEEE TRANSACTIONS ON MULTIMEDIA, IEEE MULTIMEDIA, IEEE NETWORK, IEEE WIRELESS COMMUNICATIONS, IEEE ACCESS, Journal of Network and Computer Applications (Elsevier), and International Journal of Multimedia Tools and Applications (Springer). He is a senior member ACM\end{IEEEbiography}







\end{document}